\documentclass[10pt,twocolumn,letterpaper]{article}

\usepackage{cvpr}
\usepackage{times}
\usepackage{epsfig}
\usepackage{graphicx}
\usepackage{amsmath}
\usepackage{amssymb}
\usepackage{booktabs}
\usepackage{multirow}
\usepackage{multicol}

\usepackage{bbding}

\usepackage[pagebackref=true,breaklinks=true,letterpaper=true,colorlinks,bookmarks=false]{hyperref}

\cvprfinalcopy 

\def\cvprPaperID{6079} 

\ifcvprfinal\fi
\begin{document}

\title{Medical Dataset Collection for Artificial Intelligence-based \\ Medical Image Analysis  \\
\textcolor[rgb]{1,0,0}{-- License: For Non-Commercial Use Only --}
}

\author{Yang WEN\\
School of Computer Science, UESTC\\
{\tt\small young.wen@foxmail.com}
}

\maketitle

\section{Introduction}
\label{sec: datasets}

Since finding proper dataset with sufficient well-annotated samples is vital for current AI-based research in the field of medical image analysis, I offer here a dataset collection to the community to help beginning researchers obtain the ideal dataset in a simple and straightforward way.

In total 32 public datasets was collected, of which 28 for medical imaging and 4 for natural image ones. The images of these datasets are captured by different cameras, thus vary from each other in modality, frame size and capacity. Detail information is demonstrated in Table. \ref{tab: dataset statistics}. For data accessibility, we also provide the websites of most datasets and hope this will help the readers reach the datasets.

For more information, please kindly visit my personal homepage: \url{https://wenyanger.github.io/}

\begin{table*}[t]
  \centering
  \caption{Statistics of the datasets. Here CLS denotes classification, SEG denotes segmentation, Obj-D denotes object detection, SP denotes survival prediction. For datasets with official division of training and testing sets, the number of images are shown in (number of training images) / (number of testing images), otherwise the total amounts of images.}
  \footnotesize{
    \setlength{\tabcolsep}{3.0mm}{
    \begin{tabular}{lllllll}
    \toprule
    Name  & Modality & Target & Frame Size & \# of Images & \# of Classes & Task \\
    \midrule
    CheXpert \cite{IrvinCheXpert} & X-Ray & Lung  &  $\sim$390$\times$320  & 224,316/624 & 2 & CLS \\
    ChestXRay 2017 \cite{ChestXRay_OCT_2017_kermany2018identifying} & X-Ray & Lung  &  $\sim$1000$\times$700  & 5,232/624 & 2 & CLS \\    
    LUNA \cite{luna_kaggle} & CT & Lung   & 512$\times$512 & 267  & 2 & SEG \\
    NLST~\cite{national2011national_NLST}  & CT & Lung  & -    &  - & - & SP \\
    CHAOS-CT \cite{CHAOSdata2019} & CT & Liver &   512$\times$512  & 2874 & 2 & SEG \\
    NIH-CT-82 \cite{NIH82_roth2015deeporgan}  & CT    & Pancreas &    512$\times$512 & 7,141 & 2 & SEG \\

    CHAOS-MRI \cite{CHAOSdata2019} & MRI & Liver/Kidney/Spleen &   320$\times$320  & 992 & 5 & SEG \\
    CardiacMRI \cite{cardiac_MRI_andreopoulos2008efficient} & MRI   & Heart &   256$\times$256  & 399 & 3 & SEG \\
    ACDC \cite{bernard2018deep_acdc} & MRI   & Heart &   320$\times$320  & - & 4 & SEG \\
    TCIA \cite{bloch2015data_prostate_mri_1,prostate_mri_2} & MRI   & Prostate &   320$\times$320  & - & 3 & SEG \\
    PROMISE12 \cite{litjens2014evaluation_PROMISE2012} & MRI   & Prostate &   -  & - & - & SEG \\

    ISIC2019 \cite{isic_miccai}  &  Dermoscopy     & Skin  & 1024$\times$768   &  25,331  & 9  & CLS \\

    TCGA-GBM~\cite{kandoth2013mutational_TCGA,mobadersany2018predicting_survival_2}  & H/E Stained & Cell  & -    &  - & - & SP \\
    TCGA-LGG~\cite{kandoth2013mutational_TCGA,mobadersany2018predicting_survival_2}  & H/E Stained & Cell  & -    &  - & - & SP \\

    TNBC \cite{TNBC_naylor2018segmentation}  & H/E Stained & Cell  & 512$\times$512    &  50 & 2 & SEG \\
    GlaS \cite{GlaS_sirinukunwattana2015stochastic}  & H/E Stained & Cell  & 775$\times$522  & 85/80   &  2 & SEG \\
    MoNu \cite{MoNu_kumar2017a}  & H/E Stained & Cell  & 1000$\times$1000  & 30/14   &  2 & SEG \\
    BC-2015 \cite{araujo2017classification} & H/E Stained & Breast
    cancer  & 2040$\times$1536  & 269   &  2 & CLS \\
    \midrule
    DRISHTI-GS \cite{Drishti_Sivaswamy2014Drishti} & Fundoscopic & Optic Disc, Cup & 2045$\times$1752    &  50/51 & 2 & SEG \\
    DRIONS-DB \cite{Drisons_Carmona2008Identification} & Fundoscopic & Fundus &  400$\times$400 &   & 2 & SEG \\
    RIM-r3 \cite{rimone_fumero2011rim} & Fundoscopic & Fundus &  1072$\times$1072   &  99/60 & 2 & SEG/CLS \\
    REFUGE \cite{REFUGE_orlando2020refuge} & Fundoscopic & Optic Disc, Cup/Glaucoma & 1634$\times$1634 & 400/400  & 2/2 & SEG/CLS \\
    HRF \cite{HRF_budai2013robust}  & Fundoscopic & Vessel/Glaucoma &  3504$\times$2336   &  45 & 3 & SEG/CLS \\
    DRIVE \cite{Drive_dataset} & Fundoscopic & Vessel &  565$\times$584 & 20/20 & 2 & SEG \\
    MESSIDOR \cite{messidor_decenciere2014feedback} & Fundoscopic & Diabetic Retinopathy & 2240$\times$1488 & 1,200 & 4/4 & CLS \\
    STARE \cite{STARE_dataset}  & Fundoscopic & Diabetic Retinopathy/Vessel & 700$\times$650 & 397  & 15 & SEG/CLS \\
    EyePACS \cite{EyePacs_dataset}  & Fundoscopic & Diabetic Retinopathy & 4000$\times$4000 & 35126  & 5 & CLS \\  
    \midrule  
    BUS \cite{yap2017_Ultrasound_BUS}  & Ultrasound & Breast lesion & 500$\times$500 & 163  & 2 & SEG \\  
    BUSI \cite{al2020dataset_Ultrasound_BUSI}  & Ultrasound & Breast lesion & 500$\times$500 & 780  & 3 & SEG/CLS \\  
    \midrule    
    OCT2017 \cite{ChestXRay_OCT_2017_kermany2018identifying}  & OCT   & Fundus &  $\sim$512$\times$512 & 108,309/1,000 &  4 & CLS \\
    KIMCCS \cite{IMCCS_dataset}  & Cervoscope   & Cervical &  $\sim$3000$\times$3000 & 1,463 &  3 & CLS \\
    \midrule
    CIFAR100 \cite{cifar_krizhevsky2009learning} & Natural Image & -      & 32$\times$32 &   50,000/10,000 &  100  & CLS \\
    ImageNet \cite{ilsvrc2012_krizhevsky2012imagenet} & Natural Image & -    &  $\sim$500$\times$375 & 1.3M/60000  &  1000 & CLS \\      
    ADE20K \cite{ADE20K_zhou2016semantic} & Natural Image & -    & $\sim$1000$\times$1000  & 20,210/2000 & 2603/826  & SEG/CLS \\
    PascalVOC \cite{pascalvoc_everingham2015pascal} & Natural Image & -    & $\sim$1000$\times$1000  & 20,210/2000 & 2603/826  & Obj-D \\
    \bottomrule
    \end{tabular}%
    }}
  \label{tab: dataset statistics}%
\end{table*}%

\subsection{Natural Images}
\textbf{ImageNet \cite{ilsvrc2012_krizhevsky2012imagenet}} Large Scale Visual Recognition Challenge 2012 (ILSVRC2012) is aiming to estimate the content of photographs for the purpose of retrieval and automatic annotation using a subset of the large hand-labeled ImageNet dataset (10,000,000 labeled images depicting 10,000+ object categories) as training. The general goal is to identify the main objects present in images. The training data, the subset of ImageNet containing the 1000 categories and 1.2 million images. A random subset of 50,000 of the images with labels will be released as validation data included in the development kit along with a list of the 1000 categories. The dataset is available at \textcolor[rgb]{ .753,  0,  0}{http://image-net.org/challenges/LSVRC/2012/}

\textbf{CIFAR100 \cite{cifar_krizhevsky2009learning}} The CIFAR-100 is a dataset consists of 60000 32$\times$32 colour images in 100 classes containing 600 images each. There are 500 training images and 100 testing images per class, which are randomly-selected images from each class. The 100 classes in the CIFAR-100 are grouped into 20 superclasses. In our experiments, we use the "fine" label, \ie, 100 classes, for evalution. The dataset is available at \textcolor[rgb]{ .753,  0,  0}{http://www.cs.toronto.edu/~kriz/cifar.html}

\textbf{ADE20K \cite{ADE20K_zhou2016semantic}} We use both classification and segmentation labels. The dataset is available at \textcolor[rgb]{ .753,  0,  0}{http://groups.csail.mit.edu/vision/datasets/ADE20K/}

\textbf{PascalVOC \cite{pascalvoc_everingham2015pascal}} The dataset is available at \textcolor[rgb]{ .753,  0,  0}{http://host.robots.ox.ac.uk/pascal/VOC/}


\subsection{X-Ray Images}

\textbf{CheXpert \cite{IrvinCheXpert}} CheXpert is a large public dataset for chest radiograph interpretation, consisting of 224,316 chest radiographs of 65,240 patients. The data were collected from the chest radiographic examinations from Stanford Hospital, performed between October 2002 and July 2017 in both inpatient and outpatient centers, along with their associated radiology reports. The dataset is available at \textcolor[rgb]{ .753,  0,  0}{https://stanfordmlgroup.github.io/competitions/chexpert/}

\textbf{ChestXRay 2017 \cite{ChestXRay_OCT_2017_kermany2018identifying}} The dataset was collected and labeled a total of 5,232 chest X-ray images
from children, including 3,883 characterized as depicting pneumonia (2,538 bacterial and 1,345 viral) and 1,349 normal, from
a total of 5,856 patients to train the AI system. The model was
then tested with 234 normal images and 390 pneumonia images
(242 bacterial and 148 viral) from 624 patients. The dataset is available at \textcolor[rgb]{ .753,  0,  0}{https://data.mendeley.com/datasets/rscbjbr9sj/3}

\textbf{LUNA \cite{luna_kaggle}} LUng Nodule Analysis (LUNA) is a competition held to aid the development of the nodule detection algorithm. We use the lung segmentation of the challenge for our experiments. The dataset is available at \textcolor[rgb]{ .753,  0,  0}{https://luna16.grand-challenge.org/Data/}

\textbf{MURA \cite{Rajpurkar2017MURA}} MURA (musculoskeletal radiographs) is a large dataset of bone X-rays. Algorithms are tasked with determining whether an X-ray study is normal or abnormal. The dataset is available at \textcolor[rgb]{ .753,  0,  0}{https://stanfordmlgroup.github.io/competitions/mura/}

\subsection{Ultrasound}

\textbf{BUS \cite{yap2017_Ultrasound_BUS}} The goal of the Breast Ultrasound Lesions Dataset (Dataset B) is to provide the images and ground truth for the automatic interpretation and analysis of breast ultrasound images. UDIAT-Centre Diagnostic, Corporaci Parc Taul, Sabadell (Spain) has copyright on the data and is the principal distributor of Dataset B. University of Girona and Manchester Metropolitan University are involved in an ongoing effort to develop this dataset to aid research efforts in the general area of developing, testing and evaluating algorithms for breast ultrasound lesions analysis. The dataset is available at \textcolor[rgb]{ .753,  0,  0}{www2.docm.mmu.ac.uk/STAFF/M.Yap/dataset.php}

\textbf{BUSI \cite{al2020dataset_Ultrasound_BUSI}} The data collected at baseline include breast ultrasound images among women in ages between 25 and 75 years old. This data was collected in 2018. The number of patients is 600 female patients. The dataset consists of 780 images with an average image size of 500$\times$500 pixels. The images are in PNG format. The ground truth images are presented with original images. The images are categorized into three classes, which are normal, benign, and malignant. The dataset is available at \textcolor[rgb]{ .753,  0,  0}{https://scholar.cu.edu.eg/?q=afahmy/pages/dataset}

\subsection{MRI \& CT Images}

\textbf{CHAOS \cite{CHAOSdata2019}} CHAOS challenge aims the segmentation of abdominal organs (liver, kidneys and spleen) from CT and MRI data. Segmentation of liver from computed tomography (CT) data sets, which are acquired at portal phase after contrast agent injection for pre-evaluation of living donated liver transplantation donors. Segmentation of four abdominal organs (\ie liver, spleen, right and left kidneys) from magnetic resonance imaging (MRI) data sets acquired with two different sequences (T1-DUAL and T2-SPIR). The dataset is available at \textcolor[rgb]{ .753,  0,  0}{https://chaos.grand-challenge.org/Data/}

\textbf{NIH-CT-82 \cite{NIH82_roth2015deeporgan}} The National Institutes of Health Clinical Center performed 82 abdominal contrast enhanced 3D CT scans from 53 male and 27 female subjects.  Seventeen of the subjects are healthy kidney donors scanned prior to nephrectomy.  The remaining 65 patients were selected by a radiologist from patients who neither had major abdominal pathologies nor pancreatic cancer lesions. A medical student manually performed slice-by-slice segmentations of the pancreas as ground-truth and these were verified/modified by an experienced radiologist. The dataset is available at \textcolor[rgb]{ .753,  0,  0}{https://wiki.cancerimagingarchive.net/display/Public/Pancreas-CT}

\textbf{CardiacMRI \cite{cardiac_MRI_andreopoulos2008efficient}} The dataset was comprised of short axis cardiac MR image sequences acquired from 33 subjects, for a total of 7,980 2D images. Each patient’s image sequence consisted of exactly 20 frames and the number of slices acquired along the long axis of the subjects ranged between 8 and 15. Spacing-between slices ranged between 6 and 13 mm. Each image where both the endocardial and epicardial contours of the left ventricle were visible, was manually segmented, to provide the ground truth. The manual segmentation was performed by the first author and took approx- imately 3 weeks of full time work. This resulted in 5011 manually segmented MR images, with a total of 10,022 endocardial and epicardial contours.

\textbf{ACDC \cite{bernard2018deep_acdc}} Description from~\cite{chu2020pay}:The ACDC dataset~\cite{bernard2018deep_acdc} was acquired from two MRI scanners (1.5T and 3.0T) and used in the STACOM 2017 ACDC challenge. It comprises 150 patients divided into the same 5 subgroups and annotated with three classes, namely, the right ventricles (RV), left ventricles (LV), and myocardium (Myo). The MRI data with associated annotations for 100 patients is publicly released, and the MRI data for the other 50 patients is released with their annotations hold by the challenge organizer. We leverage the MRI data for the 100 released patients with annotations for cross-dataset validation. The dataset is available at \textcolor[rgb]{ .753,  0,  0}{https://www.creatis.insa-lyon.fr/Challenge/acdc/index.html}

\textbf{Prostate segmentation} The Prostate dataset contains 40 patients from the PROSTATE-DIAGNOSIS collection~\cite{bloch2015data_prostate_mri_1} scanned with a 1.5T Philips Achieva MRI scanner. It is split into 30 patients for training, 5 for testing, and 5 for the competition (not used in our experiments). The labels are provided by Cancer Imaging Archive (TCIA) site~\cite{clark2013cancer_tcia}. The image size is 400$\times$400 or 320$\times$320. This dataset has two labeled categories, peripheral zone (PZ) and central gland (CG). The dataset is available at \href{https://wiki.cancerimagingarchive.net/display/DOI/NCI-ISBI+2013+Challenge\%3A+Automated+Segmentation+of+Prostate+Structures}{\textcolor[rgb]{ .753,  0,  0}{here}}.

\textbf{PROMISE12 Challenge \cite{litjens2014evaluation_PROMISE2012}} The MICCAI Prostate MR Image Segmentation-challenge 2012, to segment the prostate in transversal T2-weighted MR images. The data includes both patients with benign disease (e.g. benign prostatic hyperplasia) and prostate cancer. Additionaly, to test the robustness and generalizability of the algorithms, data will be from multiple centers and multiple MRI device vendors. Differences in scanning protocols will also be present in the data, e.g. patient with and without an endorectal coil. The dataset is available at \textcolor[rgb]{ .753,  0,  0}{https://promise12.grand-challenge.org/Details/}

There are 50 training cases available for download. These cases include a transversal T2-weighted MR image of the prostate. The training set is a representative set of the types of MR images acquired in a clinical setting. The data is multi-center and multi-vendor and has different acquistion protocols (e.g. differences in slice thickness, with/without endorectal coil). The set is selected such that there is a spread in prostate sizes and appearance. For each of the cases in the training set, a reference segmentation is also included.

\textbf{Why do we want to segment the prostate on MR images?}
Determination of prostate volume (PV) facilitates an assessment of prostate disorders and, for prostate cancer, in conjunction with other parameters, can help predict the pathologic stage of disease, offers insights into the prognosis, and helps predict treatment response. Prostate-specific antigen (PSA) levels have been modified to derive the PSA density by incorporating PV calculations to help guide clinical decisions. The clinical value of prostate-specific antigen density, however, is dependent on the quality of the PV estimate. The accuracy and variability of PV determinations pose limitations to its usefulness in clinical practice. Information on the size/PV, shape, and location of the prostate relative to adjacent organs is also an essential part of surgical planning for prostatectomy, radiation therapy, and emerging minimally invasive therapies, such as cryotherapy and high-intensity focused ultrasound (HIFU). Recently, the high spatial resolution and soft-tissue contrast offered by MRI makes it the most accurate method available for obtaining this kind of information. This, combined with the potential of MRI to localize and grade prostate cancer, has led to a rapid increase in its adoption and increasing research interest in its use for this application. Furthermore, and of particular relevance to the MICCAI community, is the fact that accurate prostate MRI segmentation is an essential pre-processing task for computer-aided detection and diagnostic algorithms, as well as a number of multi-modality image registration algorithms, which aim to enable MRI-derived information on anatomy and tumor location and extent to aid therapy planning and guidance.

\subsection{Hematoxylin \& Eosin Stained Images}
\noindent\textbf{Pathology Datasets Collection from Andrew Janowczyk} This page is a collection of some of my open-sourced deep learning work’s supplemental materials (i.e., tutorials  / code / datasets from papers) \textcolor[rgb]{ .753,  0,  0}{http://www.andrewjanowczyk.com/deep-learning/} \\

\noindent\textbf{A Great Review of Pathology Datasets: Machine learning methods for histopathological image analysis} \textcolor[rgb]{ .753,  0,  0}{https://arxiv.org/pdf/1709.00786.pdf} \\

\textbf{GlaS \cite{GlaS_sirinukunwattana2015stochastic}} Glands are important histological structures, which has been shown that malignant tumours arising from glandular epithelium, also known as adenocarcinomas, are the most prevalent form of cancer. Accurate segmentation of glands is often a crucial step to obtain reliable morphological statistics. Nonetheless, the task by nature is very challenging due to the great variation of glandular morphology in different histologic grades. Up until now, the majority of studies focus on gland segmentation in healthy or benign samples, but rarely on intermediate or high grade cancer, and quite often, they are optimised to specific datasets. In GlaS challenge, participants are encouraged to run their gland segmentation algorithms on images of Hematoxylin and Eosin (H\&E) stained slides. The dataset is provided together with ground truth annotations by expert pathologists. The dataset is available at \href{https://warwick.ac.uk/fac/sci/dcs/research/tia/glascontest/about/}{\textcolor[rgb]{ .753,  0,  0}{[here]}}.

\textbf{TCGA-GBM/LGG \cite{kandoth2013mutational_TCGA,mobadersany2018predicting_survival_2}} We focus on brain cancer in our study and used two public cancer survival datasets with high resolution whole slide pathological images (WSIs) from The Cancer Genome Atlas (TCGA)~\cite{kandoth2013mutational_TCGA}. Specifically, we conducted experiemnts on two cancer subtypes of brain cancer in TCGA projects: Lower-Grade Glioma (LGG) and Glioblastoma (GBM). We adopted the same annotations of vital status and overal survival time from previous study~\cite{mobadersany2018predicting_survival_2}. The TCGA-GBM dataset is available at \href{https://portal.gdc.cancer.gov/repository?facetTab=files&filters=%7B%22op%22%3A%22and%22%2C%22content%22%3A%5B%7B%22op%22%3A%22in%22%2C%22content%22%3A%7B%22field%22%3A%22cases.project.project_id%22%2C%22value%22%3A%5B%22TCGA-GBM%22%5D%7D%7D%2C%7B%22op%22%3A%22in%22%2C%22content%22%3A%7B%22field%22%3A%22files.data_type%22%2C%22value%22%3A%5B%22Slide%20Image%22%5D%7D%7D%5D%7D&searchTableTab=files}{\textcolor[rgb]{ .753,  0,  0}{[here]}}. The TCGA-LGG dataset is available at \href{https://portal.gdc.cancer.gov/repository?facetTab=files&filters=%7B%22op%22%3A%22and%22%2C%22content%22%3A%5B%7B%22op%22%3A%22in%22%2C%22content%22%3A%7B%22field%22%3A%22cases.project.project_id%22%2C%22value%22%3A%5B%22TCGA-LGG%22%5D%7D%7D%2C%7B%22op%22%3A%22in%22%2C%22content%22%3A%7B%22field%22%3A%22files.data_type%22%2C%22value%22%3A%5B%22Slide%20Image%22%5D%7D%7D%5D%7D&searchTableTab=files}{\textcolor[rgb]{ .753,  0,  0}{[here]}}. The pre-processing instruction of WSIs (in Chinese) can be found in my personal blog at \href{https://blog.csdn.net/leyounger/article/details/113775414}{\textcolor[rgb]{ .753,  0,  0}{[blog page]}}.

\textbf{NLST \cite{national2011national_NLST}} The National Lung Screening Trial (NLST) was a randomized controlled trial to determine whether screening for lung cancer with low-dose helical computed tomography (CT) reduces mortality from lung cancer in high-risk individuals relative to screening with chest radiography. Approximately 54,000 participants were enrolled between August 2002 and April 2004. The dataset is available at \textcolor[rgb]{ .753,  0,  0}{https://cdas.cancer.gov/nlst/}



\subsection{Fundoscopic \& OCT Images}

\textbf{EyePACS \cite{EyePacs_dataset}} This datasets is released along a competition on Kaggle, of which sponsored by EyePACS platform, aiming for comprehensive and automated method of diabetic retinopathy screening using image classification, pattern recognition, and machine learning. The dataset is available at \textcolor[rgb]{ .753,  0,  0}{https://www.kaggle.com/c/diabetic-retinopathy-detection/}

\textbf{DRISHTI-GS \cite{Drishti_Sivaswamy2014Drishti}} Drishti-GS is a dataset meant for validation of segmenting OD, cup and detecting notching.  The images in the Drishti-GS dataset have been collected and annotated by Aravind Eye hospital, Madurai, India. The dataset is divided into two: a training set and a testing set of images. Training images (50) are provided with groundtruths for OD and Cup segmentation and notching information. The dataset is available at \textcolor[rgb]{ .753,  0,  0}{https://cvit.iiit.ac.in/projects/mip/drishti-gs/mip-dataset2/Home.php}

\textbf{REFUGE \cite{REFUGE_orlando2020refuge}} REFUGE challenge is partnering with OMIA to widen the opportunities to present your work at MICCAI. In addition to the traditional oral and poster presentations of OMIA, REFUGE offers the chance to try your software on a real challenge this year with fundus images. The goal of the challenge is to evaluate and compare automated algorithms for glaucoma detection and optic disc/cup segmentation on a standard dataset of retinal fundus images. The dataset is available at \textcolor[rgb]{ .753,  0,  0}{https://refuge.grand-challenge.org/Home2020/}



\textbf{DRIVE \cite{Drive_dataset}} The Digital Retinal Images for Vessel Extraction (DRIVE) database has been established to enable comparative studies on segmentation of blood vessels in retinal images. The photographs for the DRIVE database were obtained from a diabetic retinopathy screening program in The Netherlands. The set of 40 images has been divided into a training and a test set, both containing 20 images. All human observers that manually segmented the vasculature were instructed and trained by an experienced ophthalmologist. They were asked to mark all pixels for which they were for at least 70\% certain that they were vessel. The dataset is available at \textcolor[rgb]{ .753,  0,  0}{https://drive.grand-challenge.org/}

\textbf{HRF \cite{HRF_budai2013robust}} High-Resolution Fundus (HRF) Image Database has been established by a collaborative research group to support comparative studies on automatic segmentation algorithms on retinal fundus images. The public database contains at the moment 15 images of healthy patients, 15 images of patients with diabetic retinopathy and 15 images of glaucomatous patients, along with the segmentation annotations of vessels. The dataset is available at \textcolor[rgb]{ .753,  0,  0}{https://www5.cs.fau.de/research/data/fundus-images/}

\textbf{MESSIDOR \cite{messidor_decenciere2014feedback}} MESSIDOR stands for Methods to Evaluate Segmentation and Indexing Techniques in the field of Retinal Ophthalmology (in French). The Messidor database has been established to facilitate studies on computer-assisted diagnoses of diabetic retinopathy. The dataset is available at \textcolor[rgb]{ .753,  0,  0}{http://www.adcis.net/en/third-party/messidor/} 

\textbf{STARE \cite{STARE_dataset}} The STARE (STructured Analysis of the Retina) Project was conceived and initiated in 1975 by Michael Goldbaum, M.D., at the University of California, San Diego. It contains ~400 images attached with corresponding diagnosis, including (0) Normal; (1) Hollenhorst Emboli; (2) Branch Retinal Artery Occlusion; (3) Cilio-Retinal Artery Occlusion; (4) Branch Retinal Vein Occlusion; (5) Central Retinal Vein Occlusion; (6) Hemi-Central Retinal Vein Occlusion; (7) Background Diabetic Retinopathy; (8) Proliferative Diabetic Retinopathy; (9) Arteriosclerotic Retinopathy; (10) Hypertensive Retinopathy; (11) Coat's; (12) Macroaneurism; (13) Choroidal Neovascularization; (14) Others. The STARE also provides some segmentation annotations of vessels. The dataset is available at \textcolor[rgb]{ .753,  0,  0}{http://cecas.clemson.edu/~ahoover/stare/}

\textbf{OCT 2017 \cite{ChestXRay_OCT_2017_kermany2018identifying}} The dataset obtained 108,312 images
(among these, there are 37,206 with choroidal neovascularization, 11,349 with diabetic
macular edema, 8,617 with drusen, and 51,140 normal) from
4,686 patients. The dataset is available at \textcolor[rgb]{ .753,  0,  0}{https://data.mendeley.com/datasets/rscbjbr9sj/3}

\subsection{Dermoscopy Images}
\textbf{ISIC 2019 \cite{isic_miccai,isic2_tschandl2018the,isic3_combalia2019bcn20000:}} The International Skin Imaging Collaboration (ISIC) has developed the ISIC Archive, an international repository of dermoscopic images, for both the purposes of clinical training, and for supporting technical research toward automated algorithmic analysis for skin cancer. The goal for ISIC 2019 is classify dermoscopic images among nine categories: 1. Melanoma; 2. Melanocytic nevus; 3. Basal cell carcinoma; 4. Actinic keratosis; 5. Benign keratosis (solar lentigo / seborrheic keratosis / lichen planus-like keratosis); 6. Dermatofibroma; 7. Vascular lesion; 8. Squamous cell carcinoma; 9. None of the others. There are 25,331 images available for training across 8 different categories. The dataset is available at \textcolor[rgb]{ .753,  0,  0}{https://challenge2019.isic-archive.com/}

{\small
\bibliographystyle{ieee_fullname}
\bibliography{cvpr2020}
}

\end{document}